\documentclass[12pt]{article}

\usepackage{latexsym}
\usepackage{color}

\long\def\nop#1{}

\def\comment{\edef\cps{\the\parskip} \parskip=0.5cm \begingroup \tt}

\expandafter\ifx\csname shortcite\endcsname\relax

\fi

\newbox\current

\long\def\plframebox#1{
\setbox\current\vbox{#1}		

\vbox to \ht\current {\hrule\vss
\hbox to \wd\current {%
\vrule \hss\box\current\hss \vrule}
\vss\hrule }
}

\long\def\eatpar#1{%
\ifx#1\par                      
\let\nextmove=\eatpar           
\else
\let\nextmove=#1
\fi
\noexpand\nextmove
}

\def\modifymargins#1#2{
\newdimen\addtoh
\newdimen\addtow
\addtoh=#1
\addtow=#2

\advance\topmargin by -\addtoh
\multiply\addtoh by 2
\advance\textheight by \addtoh

\advance\oddsidemargin by -\addtow
\advance\evensidemargin by -\addtow
\multiply\addtow by 2
\advance\textwidth by \addtow
}

\begingroup
\catcode`\~=11
\gdef\centertilde#1{\lower #1pt\hbox{~}}
\endgroup

\newcount\currenttime
\newcount\hour
\newcount\minute

\def\printtime{%
\currenttime=\time
\hour=\currenttime
\divide\hour by 60
\minute=-\hour
\multiply\minute by 60
\advance\minute by \currenttime
\the\hour:\ifnum\minute<10 0\fi\the\minute
}

\begingroup
\makeatletter
\global\let\@@date=\@date
\gdef\@date{\@@date\ --- \printtime}
\endgroup

\def\oggi{\number\day\space 
\ifcase\month\or
Gennaio\or Febbraio\or Marzo\or Aprile\or Maggio\or Giugno\or
Luglio\or Agosto\or Settembre\or Ottobre\or Novembre\or Dicembre\fi
\space \number\year}

\newcounter{rmexample}

\def\proof{\noindent {\sl Proof.\ \ }}

\def\qed{\hfill{\boxit{}}
  \ifdim\lastskip<\medskipamount \removelastskip\penalty55\medskip\fi}
\def\qedn#1{\hfill{\boxit{}$_#1$}
  \ifdim\lastskip<\medskipamount \removelastskip\penalty55\medskip\fi}
\long\def\boxit#1{\vbox{\hrule\hbox{\vrule\kern3pt
                  \vbox{\kern3pt#1\kern3pt}\kern3pt\vrule}\hrule}}

  \def\G{{\cal G}} 
\def\I{{\cal I}}   
\def\M{{\cal M}} \def\N{{\cal N}} \def\O{{\cal O}}

\def\ie{i.e.}
\def\eg{e.g.}

\def\M{{\cal M}}

\def\l{\langle}
\def\r{\rangle}

\def\mod{M\!od}

\def\var{V\!ar}

\def\p{{\rm P}}
\def\np{{\rm NP}}

\def\S#1{\mbox{$\Sigma^p_{#1}$}}
\def\P#1{\mbox{$\Pi^p_{#1}$}}

\def\nuc#1{\mbox{$\parallel\!\leadsto$#1}}

\def\nucp{\nuc{\rm P}}
\def\nucnp{\nuc{\rm NP}}

\def\profont{\sf}

\def\sat{{\profont sat}}

\def\threesat{{\profont 3sat}}

\def\x3c{{\profont x3c}}

\def\strips{{\profont STRIPS}}

\def\possnewtheorem#1#2{
\expandafter\ifx\csname #1\endcsname\relax
\newtheorem{#1}{#2}
\fi
}

\def\possnewtheoremthree#1[#2]#3{
\expandafter\ifx\csname #1\endcsname\relax
\newtheorem{#1}[#2]{#3}
\fi
}

\possnewtheorem{theorem}{Theorem}
\possnewtheorem{corollary}{Corollary}
\possnewtheorem{lemma}{Lemma}
\possnewtheoremthree{proposition}[theorem]{Proposition}
\possnewtheorem{definition}{Definition}
\possnewtheorem{question}{Question}
\possnewtheorem{example}{Example}
\possnewtheorem{nontheorem}{Counterexample}
\possnewtheorem{property}{Property}
\possnewtheorem{assumption}{Assumption}
\possnewtheorem{conjecture}{Conjecture}
\newenvironment{theorem*}[1]{{\noindent \bf Theorem~#1}\begin{it}}{\end{it}\

}

\modifymargins{0pt}{10pt}

\def\comment{\vskip 0.5cm \begingroup \tt}

\def\plansat{{\profont PLANSAT}}

\title{The Complexity of Modified Instances}
\author{Paolo Liberatore}
\date{}

\begin{document}

\maketitle

\begin{abstract}

In this paper we study the complexity of solving a problem
when a solution of a similar instance is known. This problem
is relevant whenever instances may change from time to time,
and known solutions may not remain valid after the change.
We consider two scenarios: in the first one, what is known
is only a solution of the problem before the change; in the
second case, we assume that some additional information,
found during the search for this solution, is also known. In
the first setting, the techniques from the theory of
NP-completeness suffice to show complexity results. In the
second case, negative results can only be proved using the
techniques of compilability, and are often related to the
size of considered changes.

\end{abstract}

 %

\section{Introduction}

Many complexity results in the literature are about simple
decision problems, such as: is a propositional formula
consistent? is there a vertex cover of a graph containing at
most $k$ nodes? is there a plan to achieve this goal?  etc.
These problems are often simplification of real-world
problems. In practice, what is needed is not only a ``yes/no''
answer. For instance, from the problem of planning what is
really needed is a plan to achieve the goal, not only an
answer to the question of existence.

Another property of these simplified problem is the absence
of any hint in their specification. Constraints and hints
are different:
\eatpar

\begin{description}

\item[constraint:] affects the specification of the problem:
if a constraint is added or removed, the set of solutions of the
problem may change;

\item[hint:] something that may help finding the solution; it
can be neglected without changing the set of solutions.

\end{description}

In this paper we consider some problems in which both
constraints and hints are present. Namely, if the available
data is composed of an instance of a problem, and a solution
for a similar instance, we have both constraints (the
instance) and an hint (the solution of the similar
instance). This scenario may occur in practice, \eg, a
solution for a specific instance is known, and then the
instance is changed. In the planning domain, for example, it
may be that, during the execution of a plan, some conditions
that were previously assumed to hold, are indeed false, and
some actions in the plan cannot be performed any more. In
such a case, a new plan (from a different initial state) is
needed. In this case, besides the instance specification
(the description of initial state, goal, and actions) an
hint is also avaliable: the plan for the previous instance.
In some cases, this may be of help, because some parts of
the plan can be reused for the modified instance.

A similar definition can be given for almost all search problem,
although in some cases the analysis is only of theoretical
interest. For instance, we can consider what happens if we want
to solve the satisfiability problem when a solution for a
similar instance is known, but this problem does not seem to
have a practical relevance. We study it nevertheless, as it its
analysis is very simple to carry on, allowing for a clear
explanation of definitions and methods used for other problems
for which changes are of more practical interest.

Besides satisfiability and planning, we also consider the
problem of vertex cover. Namely, given an optimal vertex cover
for a graph, we consider the problem of finding a vertex cover
for a graph obtained by adding one or more edges to the
previous graph.

In all those cases, we consider the complexity of the
corresponding decision problem. For example, for the case of
vertex cover, we want to determine the complexity of deciding
whether there exists a vertex cover composed of at most $k$
nodes, for the graph $(N,E \cup \{e\})$, if an optimal vertex
cover for $(N,E)$ is known.

The analysis is done using quite standard complexity
technique (and is in many cases very simple), that is, the
theory of \np-completeness. However, there is a natural
generalization of these problems that cannot be analyzed
using standard complexity classes. Consider the case in
which we want to find a plan, and we are already alerted
that the plan may fail because of a change in the
specification of the initial state. What we can do is to
``save'' some intermediate results which may allow for
updating the plan when a new initial state is defined.

It is quite easy to see that the classes of the polynomial
hierarchy and the polynomial many-one reduction are useless for
solving this problem. Indeed, the problem is not completely
specified, since there is no way of formalizing a concept such
as ``an intermediate result'' of an algorithm, where moreover
the algorithm itself is not specified. In practice, the problem
is ``how to reduce the problem of satisfiability to the problem
of updating a plan when some intermediate results of some
algorithm are known?''. If we consider that anything can be an
intermediate result, and any algorithm is a possible candidate,
we see that giving a polynomial reduction from \eg,
satisfiability, seems to be very hard.

This problem can be solved very easily using compilability
classes. Indeed, we can generalize further the notion of
``intermediate result'', saying that the hint we have is any
polynomial-sized data structure depending only on the original
instance.  Since compilability allows for giving theorems of
negative results in this case, these negative theorems also
extend to the case in which only intermediate results of
algorithms are considered.

 %

\section{Propositional Satisfiability}

In order to show the concepts used in the paper, we begin
analyzing the problem of solving a modified instance for the
problem of propositional satisfiability. The proofs of this
sections are quite straightforward, and the results are completely
unsurprising. This simplicity allows for an easier presentation of
the definitions and techniques used to prove our results.

\subsection{Fixed Model}
\label{fixed-model}

Let us consider a set of propositional clauses $F$, over an
alphabet $X$. Assume that a model $I \in \mod(F)$ is known. For
some reason, this set $F$ is changed, for instance a clause is
added, and we want to determine a model for this new instance $F
\cup \{ \gamma \}$. Disregarding the model $I$, finding a model of
a formula $F \cup \{ \gamma \}$ is \np-equivalent. Knowing a
model of a similar formula cannot make the problem harder, so
it is still \np: in general, any upper bound for the problem with
no hint extends to the problem with hints.

The knowledge of the model $I$ may help. For instance, if
$I \models \gamma$, then $I$ is itself a solution of
$F \cup \{ \gamma \}$. In this case, finding a model for the
modified instance only takes linear time. However, if
$I \not\models \gamma$, then $I$ may be useless for finding a
model of $F \cup \{ \gamma \}$.

In order to formally prove our intuition that a model cannot
help in general for solving a modified instance, we consider
the related decision problem, and establish its complexity.
Given a model $I$ of a set of clauses $F$, and a clause
$\gamma$, we want to decide whether $F \cup \{ \gamma \}$ is
satisfiable. Note that $F$ is satisfiable, since it has at
least a model. This problem is clearly in \np\  (disregard $I$,
and check satisfiability of $F \cup \{ \gamma \}$). This is an
{\em upper bound} of the problem. 

Finding a {\em lower bound} is slightly more complex. The data of
the problem is not only composed of a set of clauses, but rather
of a triple $\l F, \gamma, I \r$, where $F$ is a set of clauses,
$\gamma$ is another clause, and $I$ is a model of $F$. The
problem is that the third part of the input (the model $I$) does
not put any additional constraint (that is, the answer can be
determined from $F$ and $\gamma$).  Rather, it is an {\em hint},
that is, a suggestion that can be taken or not.

A proof of \np-hardness can be obtained by reduction from \sat.
We prove that there exists a polynomial reduction from the
problem of checking satisfiability of a set of clauses $G$ to the
problem under consideration. Namely, if $\l F, \gamma, I \r$ is
the instance obtained by applying the reduction to $G$, then $G$
is satisfiable if and only if $\l F, \gamma, I \r$ is a valid
instance. Validity of $\l F, \gamma, I \r$ means that $I$ is a
model of $F$, and $F \cup \{\gamma\}$ is satisfiable. Such a
reduction is very easy to give:
\eatpar

\[
G ~~ \Rightarrow ~~ \l a \vee G, \neg a, \{a\} \r
\]

\noindent where $a \vee G$ is a shorthand for $\{ a \vee \gamma
~|~ \gamma \in G\}$. Indeed, $\{a\}$ is a model of $a \vee G$,
but $(a \vee G) \cup \{\neg a\}$ is equivalent to $G \cup \{\neg
a\}$, thus it is satisfiable if and only if $G$ is satisfiable
(note that $\{a\}$ is not a model of this formula).

We have therefore proved that the problem of checking, given
$\l F, \gamma, I \r$, whether $I$ is a model of $F$, and
$F \cup \{ \gamma \}$ is satisfiable, is \np-hard (a quite
trivial result). The next subsection shows that such hardness
proof may not be suitable. This calls for a formal definition
of this kind of reductions.

We express a search problem in \np\  by a polynomial binary
relation over pairs of strings: the problem is that
of finding a string $y$ such that $\l x,y \r \in B$, given $x$.
This is the problem for which we want to check whether the
solution of an instance may help for solving modified instances.

To formalize instance changes, we use a function $c(x,y)$ that
takes two strings $x$ and $y$, and gives a new string $z$ obtained
by ``merging'' in some way $x$ and $y$. Intuitively, $c$ is the
function used to modify an instance. In the case of satisfiability,
$x$ is a set of clauses, $y$ is a clause, and $c(x,y)=x \cup \{y\}$.
For other problems, $c$ may have a different definition.

The relation $B$ and the function $c$ define the problem and the
way it is modified, so they suffice to formally define the problem
of modified instances. The hardness result used for satisfiability
is actually the \np-hardness of the following problem $C$.
\eatpar

\[
C = \{ \l x,y,z \r ~|~
	\l x,z \r \in B \mbox{ and }
	\exists z' ~.~ \l c(x,y), z' \r \in B \}
\]

Informally, $C$ is the problem whose instances are triples
$\l x, y, z \r$, and the ``yes'' instances are those such
that $z$ is a solution of $x$, and $c(x,y)$ is satisfiable
(this is the instance obtained by modifying $x$ with $y$).

In order to show the \np-hardness of this problem, we take an
already known \np-hard problem $A$, and reduce it to $C$ in
polynomial time. In other words, we find three polynomial
functions $f$, $g$, and $h$, such that:\eatpar

\[
\forall x ~.~
x \in A ~~ \Leftrightarrow ~~ \l f(x), g(x), h(x) \r \in C
\]

Summarizing, from the search problem $B$ we define a decision
problem $C$, and and then show a reduction from an \np-hard
problem $A$ to $C$, thus proving that $C$ is \np-hard.

\subsection{Arbitrary Model}
\label{arbitrary-model}

The proof given in the previous section has a drawback. Let
us consider the reduction used for \sat. What we proved is
only that, in general, there exists a model of $a \vee G$
that is useless for solving the satisfiability of $(a \vee
G) \cup \{\neg a\}$. However, there may be other, more useful,
models. Let us assume, for example, that the model
of $a \vee G$ is chosen in the following ``unfair'' way:
\eatpar

\begin{enumerate}
\itemsep=0pt

\item if $G$ is satisfiable, then $I$ is a model of $G$;

\item otherwise, $I = \{a\}$.

\end{enumerate}

Clearly, such model $I$ allows for solving the satisfiability
of $(a \vee G) \cup \{\neg a\}$ in linear time. This is not
in contradiction with the \np-hardness result. Indeed, hardness
only proves the existence of hard instances, and not that all
instances of a problem are hard. In this case, hardness shows
that there are models of the original instance that do not
help in solving the modified one.

This example shows that, while there is some model of $F$
that does not help in solving satisfiability of $F \cup
\{\gamma\}$, some other model may help. If we could prove
that, for each formula $F$ there is a model $I$ that makes
the satisfiability of $F \cup \{\gamma\}$ polynomial, we would
have proved that a careful choice of the solution of the original
instance actually helps the solution of the modified one.

In order to prove the impossibility of such method, we would
need to modify the reduction so that, for {\em any} model
of $F$, verifying satisfiability of $F \cup \{\gamma\}$ is hard.

It is easy to show that, if only one clause is added, such a
reduction does not exist. Indeed, we can always set $I$ to be a
model of $F \cup \{\gamma\}$ if it is satisfiable, and thus
there is a model that helps solving the modified instance.
This proves that proving hardness this way is impossible.
However, it does not provide a method for finding a model that
helps solving any possible modified instance, as $I$ cannot
be determined without knowing the clause $\gamma$ to be added.
As a result, we do not have an efficient solution method, and
we know we cannot prove its unfeasibility by using hardness in
the way we planned.

If we allow clauses to be either added or deleted, we can really
prove that no model of a formula helps checking the
satisfiability a modified formula. Let us assume that a formula
may be modified by adding a clause $\gamma$ and removing another
clause $\delta$. Given a model $I$ of $F$, we want to check
satisfiability of $F \cup \{\gamma\} \backslash \{\delta\}$.

A proof of \np-hardness is the following one. Let $G$ be a
propositional formula. We define $F$ to be $\{ a \} \cup ( ( X
\cup \{a\} ) \vee (G \cup \{\neg a\} ) )$, where $F \vee G$ is a
shorthand for $\{ \gamma \vee \delta ~|~ \gamma \in F ,~ \delta
\in G \}$. This formula has a single model: $\{a\} \cup X$.
Furthermore, if we set $\gamma=\{\neg a\}$ and $\delta=\{a\}$, we
have that $F \cup \{\gamma\} \backslash \{\delta\}$ is equivalent
to $G \cup \{\neg a\}$, which is satisfiable if and only if $G$ is
satisfiable.

This way, we have proved that, given {\em any} model of $F$,
checking satisfiability of $F \cup \{\gamma\} \backslash
\{\delta\}$ is \np-hard.  For a generic search problem expressed
by a binary relation $B$, the technique is to prove that there
exists an \np-hard problem $A$ and two polynomial functions $f$
and $g$ such that ($C$ is defined as in the previous section):
\eatpar

\[
\forall x,z ~.~ x \in A ~~ \Leftrightarrow ~~
( \l f(x), g(x), z \r \in C \Leftrightarrow \l f(x),z \r \in B)
\]

This proves that the problem remain hard no matter how the
solution of the original instance $z$ is chosen. In practice,
the only way for proving the claim is by giving a reduction in
such a way $f(x)$ has exactly one solution. This statement is
slightly stronger than the previous one: it cannot be proved
for \sat\  in the assumption that the only possible change is
the addition of a clause. Changes involving removing of
clauses were needed.

\subsection{Arbitrary Polynomial Data Structure}
\label{polynomial-information}

We now consider the most general possible setting. In the
analysis done so far, we assumed that what is known is only a
solution of a given instance. However, it is reasonable to
assume that some other information is given. For instance, we
may save some intermediate results obtained during the search
of the first solution, and this information may be of help in
looking for further solutions. In the case of satisfiability,
we may have information about the satisfiability of some
subformulae, etc.

In general, we may assume the the size of this piece of
information is bounded by a polynomial in the size of the
input. Since we want results that are independent from the
algorithms used, we have to make the fewer possible
assumptions. As a result, we assume that what is known is
simply a piece of information related to the formula $F$. In
other words, we are given the formula $F$, a list of changes
$D$ (additions and deletions of clauses), and some piece
of information $I$, whose size is polynomial in the size of
$F$. We want to check whether the formula obtained by
modifying $F$ with $D$ is satisfiable or not.

The case in which $D$ is composed only of a fixed number of
changes (either additions or deletions of clauses) is easy.
Indeed, if $D$ is bounded to be composed of at most $k$ changes,
the total number of possible changes is $O(2^k)$, and this is
a constant as well. Therefore, there are only a constant
number of possible changes.

A data structure $I$ that allows for the solution of instances
after the change is a table that reports, for each possible change,
the solution of the resulting instance. Note, indeed, that we
have not given any restriction on $I$, besides it must be of
polynomial size. Given $I$ and any possible change of size
bounded by $k$, finding a solution of the modified instance is
just a matter of a table lookup.

A polynomially-sized data structure is
therefore more useful than having simply a model of the original
instance, as an hint. Indeed, the former helps in finding a
solution after a constant number of changes, while the latter
does not, even for the case of two changes only (as proved in
the last section).

If the set of changes $D$ is not composed of a fixed number
of changes, we can prove that any polynomial-sized piece of
information is in general useless. This result is not
surprising, as in practice we can obtain any set of clauses
$F'$ by a number of changes over any other set of clauses $F$.
Therefore, having a data structure $I$ that depends on $F$
is equivalent of having a constant data structure, and this
cannot help in solving the general problem of satisfiability.
Formally, if the original set is $F=\emptyset$, then its size
does not depends on the size of $D$. Therefore, the size of
$I$ is unrelated to the size of $D$ as well. Now, any clause
can be expressed as the addition of a suitable number of
clauses to $\emptyset$. Considering the satisfiability of this
set of clauses, the size of $I$ is a constant. As a result,
if this problem is polynomial, then the problem of satisfiability
is polynomial as well, as it can be solved in polynomial time
using a constant (independent from the set of clauses) data
structure $I$.

On the other hand, it is clear that, if {\em any} change is
allowed, no gain can be achieved, as any possible instance
of the problem can be obtained by suitably changing a trivial
instance (like $F=\emptyset$). In practice, as for example in
the planning case that will be shown in the sequel, the kind
of possible changed are often limited to a part of the instance.
The most interesting case, in the satisfiability problem, is
that of changing a set of clauses without adding new variables.
This restrict the set of changes, as for example no changes
are possible to $F=\emptyset$. Given a set of clauses, we can
change it only if the resulting set of clauses has the same
set of variables, or a subset of the original one.

The set of possible changes is now restricted, but it nevertheless
allows for proving that, in general, a polynomial data structure
depending on a single instance will not help in solving similar
instance. Intuitively, this is due to the fact that any set of
clauses on the same set of variables can be obtained from a given
one. There are exponentially many of such sets, while the data
structure is only polynomially large. It looks therefore unlikely
that a polynomial data structure can help in solving exponentially
many instances. However, this is not a complete proof, and it is
also easy to construct problems for which useful data structures
exist.

The formal proof of impossibility for \sat\  is based on the
following problem $n\sat$:\eatpar

\[
n\sat= \{ \l x,y \r ~|~ y \in \sat \mbox{ and } x=1^{|\var(y)|} \} 
\]

In words, the first part of the input is a string whose only
information is the number of variables of the formula represented
by the varying part. It is easy to show that $n\sat$ is
\nucnp-complete~\cite{cado-etal-96,cado-etal-02}.

Adding and removing an arbitrary number of clauses from a set
$F$ we can obtain any set of clauses with the same set of variables.
This means that the existence of a data structure $I$ of polynomial
size, that makes the solution of the changed instance polynomial
is the same as the existence of an $I$, depending only on the
number of variable, that makes the solution of the satisfiability
problem for all sets of clauses over the given alphabet polynomial.

As a result, if there exists a way of determining some piece of
information $I$ such that solving any modified instance is
polynomial, then $n\sat$ would be solvable in polynomial time,
once a preprocessing step (of unbounded time length, but producing
a polynomially-sized output). This implies that $n\sat$ is
in \nucp. The membership of a \nucnp-hard problem to \nucp\ 
implies the collapse of the polynomial hierarchy to its second
level~\cite{cado-etal-96,cado-etal-02}.

\subsection{Summary}

The analysis of \sat, from the point of view of instance modification,
is quite straightforward. However, it allows for showing that different
possible formal definition of the problem are possible. In particular,
the instance modification problem is:

\begin{enumerate}

\item given an instance and a change, is there {\em at least a model}
of the instance that helps solving the modified instance?
(Subsection~\ref{fixed-model}: fixed model);

\item given an instance and a change, does {\em any model} of the
instance help in solving the modified instance?
(Subsection~\ref{arbitrary-model}: arbitrary model);

\item given an instance, and a change, is there any polynomially-sized
data structure, depending only on the instance (not on the change),
that makes the modified instance easier to solve?
(Subsection~\ref{polynomial-information}: polynomial data structure).

\end{enumerate}

For \sat, the first is impossible even if the change is the simple
addition of a unit clause; the second can be proved impossible only
allowing a unit clause to be added and another one to be deleted.
The third case depends on additional constraints on the change:

\begin{enumerate}

\item if only a constant number of additions/deletions is allowed,
then the modified instance problem becomes polynomial;

\item if any change is possible, the problem can be proved to be
as hard as the original one with a very simple proof (using only
classes of the polynomial hierarchy);

\item if changed are restricted to be about clauses on the same
alphabet of the original formula, then the modified instance
problem can be proved to be as hard as the original one, but the
proof requires the use of compilability classes.

\end{enumerate}

 %

\section{Vertex Cover}

We consider the problem of vertex cover, and assume that
the possible changes are addition of edges. The case with
also removal of edges is similar.

Let $G=(N,E)$ be a graph. We assume that an optimal vertex
cover $V$ is known, and a set of edges $E'$ is added. The
modified instance is thus $G'=(N,E \cup E')$. As in the
case of satisfiability, we are looking for an optimal
vertex cover for the new instance, but we only consider a
related decision problem: deciding whether there exists a
vertex cover for $G'$ of size at most $k$, the latter being
part of the input. Clearly, if an optimal vertex cover for
$G'$ can be found, then this decision problem can be easily
solved.

We show that there exists two polynomial functions $f$ and
$g$ such that, given a set of clauses $F$, $f(F)$ is a
graph, $g(G)$ is an integer, and it holds:\eatpar

\begin{enumerate}
\itemsep=0pt

\item $F$ is satisfiable if and only if there exists a
vertex cover of $f(F)$ of size at most $g(F)$;

\item adding or removing a unit clause to $F$ is equivalent
to adding an edge to $f(F)$ and modifying $k$ (formally, it
is possible to add an edge to $f(F)$ in such a way the
resulting graph has a vertex cover of size $g(F)$ if and
only if $F \cup \{\gamma\}$ is satisfiable, and the same
for removing of clauses, adding 1 to $k$).

\end{enumerate}

This can be proved as follows. We consider the usual
reduction from \threesat\ to vertex cover. Each variable
$x_i$ is translated into a pair of nodes joined by an edge.
One of the two nodes is associated with $x_i$, the other
with $\neg x_i$.  Each clause $\gamma_i$ composed of $j$
literals is mapped into a clique of $j$ nodes. If $\gamma_i
= \vee l_j$, we call these nodes $n^i_j$. Each node of this
clique is then joined to the node associated with the
corresponding literal. In order to allow for
adding/removing of unit clauses, we add two nodes for each
literal, say $l_i'$ and $l_i''$, even if the corresponding
unit clause is not in $F$. If the unit clause $l_i$ is in
$F$, we join $l_i$ with $l_i'$.

Below is the graph corresponding to the set of clauses
$F=\{\gamma_1, \gamma_2\}$, where $\gamma_1=x_1 \vee x_2$,
while $\gamma_2=\neg x_1$.

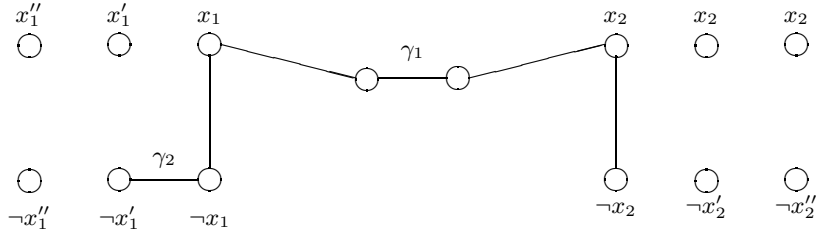
\begin{center}
\setlength{\unitlength}{3108sp}%
\begingroup\makeatletter\ifx\SetFigFont\undefined%
\gdef\SetFigFont#1#2#3#4#5{%
  \reset@font\fontsize{#1}{#2pt}%
  \fontfamily{#3}\fontseries{#4}\fontshape{#5}%
  \selectfont}%
\fi\endgroup%
\begin{picture}(6612,1834)(999,-2189)
{\color[rgb]{0,0,0}\thinlines
\put(1981,-691){\oval(180,180)}
}%
{\color[rgb]{0,0,0}\put(2701,-691){\oval(180,180)}
}%
{\color[rgb]{0,0,0}\put(1981,-1771){\oval(180,180)}
}%
{\color[rgb]{0,0,0}\put(2701,-1771){\oval(180,180)}
}%
{\color[rgb]{0,0,0}\put(3956,-971){\oval(180,180)}
}%
{\color[rgb]{0,0,0}\put(4681,-961){\oval(180,180)}
}%
{\color[rgb]{0,0,0}\put(5946,-701){\oval(180,180)}
}%
{\color[rgb]{0,0,0}\put(6666,-701){\oval(180,180)}
}%
{\color[rgb]{0,0,0}\put(5941,-1771){\oval(180,180)}
}%
{\color[rgb]{0,0,0}\put(6666,-1781){\oval(180,180)}
}%
{\color[rgb]{0,0,0}\put(1266,-701){\oval(180,180)}
}%
{\color[rgb]{0,0,0}\put(1266,-1781){\oval(180,180)}
}%
{\color[rgb]{0,0,0}\put(7381,-691){\oval(180,180)}
}%
{\color[rgb]{0,0,0}\put(7381,-1771){\oval(180,180)}
}%
{\color[rgb]{0,0,0}\put(2791,-691){\line( 4,-1){1080}}
}%
{\color[rgb]{0,0,0}\put(4051,-961){\line( 1, 0){540}}
}%
{\color[rgb]{0,0,0}\put(4771,-961){\line( 4, 1){1080}}
}%
{\color[rgb]{0,0,0}\put(2071,-1771){\line( 1, 0){540}}
}%
{\color[rgb]{0,0,0}\put(5941,-781){\line( 0,-1){900}}
}%
{\color[rgb]{0,0,0}\put(2701,-781){\line( 0,-1){900}}
}%
\put(4321,-781){\makebox(0,0)[b]{\smash{{\SetFigFont{9}{10.8}{\familydefault}{\mddefault}{\updefault}{\color[rgb]{0,0,0}$\gamma_1$}%
}}}}
\put(2701,-511){\makebox(0,0)[b]{\smash{{\SetFigFont{9}{10.8}{\familydefault}{\mddefault}{\updefault}{\color[rgb]{0,0,0}$x_1$}%
}}}}
\put(2701,-2131){\makebox(0,0)[b]{\smash{{\SetFigFont{9}{10.8}{\familydefault}{\mddefault}{\updefault}{\color[rgb]{0,0,0}$\neg x_1$}%
}}}}
\put(1981,-2131){\makebox(0,0)[b]{\smash{{\SetFigFont{9}{10.8}{\familydefault}{\mddefault}{\updefault}{\color[rgb]{0,0,0}$\neg x_1'$}%
}}}}
\put(1261,-2131){\makebox(0,0)[b]{\smash{{\SetFigFont{9}{10.8}{\familydefault}{\mddefault}{\updefault}{\color[rgb]{0,0,0}$\neg x_1''$}%
}}}}
\put(1261,-511){\makebox(0,0)[b]{\smash{{\SetFigFont{9}{10.8}{\familydefault}{\mddefault}{\updefault}{\color[rgb]{0,0,0}$x_1''$}%
}}}}
\put(1981,-511){\makebox(0,0)[b]{\smash{{\SetFigFont{9}{10.8}{\familydefault}{\mddefault}{\updefault}{\color[rgb]{0,0,0}$x_1'$}%
}}}}
\put(2341,-1636){\makebox(0,0)[b]{\smash{{\SetFigFont{9}{10.8}{\familydefault}{\mddefault}{\updefault}{\color[rgb]{0,0,0}$\gamma_2$}%
}}}}
\put(5941,-511){\makebox(0,0)[b]{\smash{{\SetFigFont{9}{10.8}{\familydefault}{\mddefault}{\updefault}{\color[rgb]{0,0,0}$x_2$}%
}}}}
\put(5941,-2041){\makebox(0,0)[b]{\smash{{\SetFigFont{9}{10.8}{\familydefault}{\mddefault}{\updefault}{\color[rgb]{0,0,0}$\neg x_2$}%
}}}}
\put(6661,-511){\makebox(0,0)[b]{\smash{{\SetFigFont{9}{10.8}{\familydefault}{\mddefault}{\updefault}{\color[rgb]{0,0,0}$x_2$}%
}}}}
\put(6661,-2041){\makebox(0,0)[b]{\smash{{\SetFigFont{9}{10.8}{\familydefault}{\mddefault}{\updefault}{\color[rgb]{0,0,0}$\neg x_2'$}%
}}}}
\put(7381,-511){\makebox(0,0)[b]{\smash{{\SetFigFont{9}{10.8}{\familydefault}{\mddefault}{\updefault}{\color[rgb]{0,0,0}$x_2$}%
}}}}
\put(7381,-2041){\makebox(0,0)[b]{\smash{{\SetFigFont{9}{10.8}{\familydefault}{\mddefault}{\updefault}{\color[rgb]{0,0,0}$\neg x_2''$}%
}}}}
\end{picture}%
 %

\end{center}

It is not difficult to prove that there exists a vertex
cover of $n+m-r$ nodes if and only if the set of clauses is
satisfiable, where $n$ is the number of variables, $m$ is
the total number of {\em occurrences of literals} and $r$
is the number of clauses.

Moreover, adding a unit clause $l_i$ is equivalent to
adding an edge joining the node $l_i$ and the node $l_i'$.
Namely, $F \cup \{l\}$ is satisfiable if and only if the
graph obtained by adding an edge between $l_i$ and $l_i'$
has a vertex cover of size $n+m-r$. Below is the graph
modified by the addition of the unit clause $\gamma_3=\neg
x_2$.

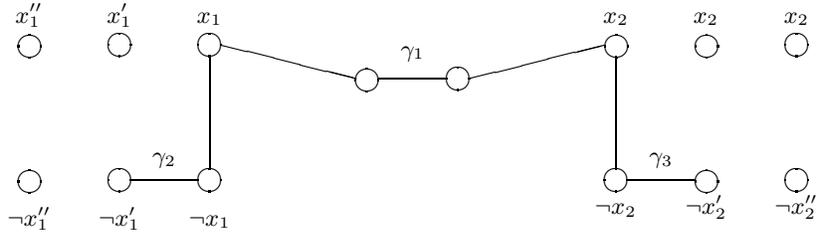
\begin{center}
\setlength{\unitlength}{3108sp}%
\begingroup\makeatletter\ifx\SetFigFont\undefined%
\gdef\SetFigFont#1#2#3#4#5{%
  \reset@font\fontsize{#1}{#2pt}%
  \fontfamily{#3}\fontseries{#4}\fontshape{#5}%
  \selectfont}%
\fi\endgroup%
\begin{picture}(6643,1834)(968,-2189)
{\color[rgb]{0,0,0}\thinlines
\put(1981,-691){\oval(180,180)}
}%
{\color[rgb]{0,0,0}\put(2701,-691){\oval(180,180)}
}%
{\color[rgb]{0,0,0}\put(1981,-1771){\oval(180,180)}
}%
{\color[rgb]{0,0,0}\put(2701,-1771){\oval(180,180)}
}%
{\color[rgb]{0,0,0}\put(3956,-971){\oval(180,180)}
}%
{\color[rgb]{0,0,0}\put(4681,-961){\oval(180,180)}
}%
{\color[rgb]{0,0,0}\put(5946,-701){\oval(180,180)}
}%
{\color[rgb]{0,0,0}\put(6666,-701){\oval(180,180)}
}%
{\color[rgb]{0,0,0}\put(5941,-1771){\oval(180,180)}
}%
{\color[rgb]{0,0,0}\put(6666,-1781){\oval(180,180)}
}%
{\color[rgb]{0,0,0}\put(1266,-701){\oval(180,180)}
}%
{\color[rgb]{0,0,0}\put(1266,-1781){\oval(180,180)}
}%
{\color[rgb]{0,0,0}\put(7381,-691){\oval(180,180)}
}%
{\color[rgb]{0,0,0}\put(7381,-1771){\oval(180,180)}
}%
{\color[rgb]{0,0,0}\put(2791,-691){\line( 4,-1){1080}}
}%
{\color[rgb]{0,0,0}\put(4051,-961){\line( 1, 0){540}}
}%
{\color[rgb]{0,0,0}\put(4771,-961){\line( 4, 1){1080}}
}%
{\color[rgb]{0,0,0}\put(2071,-1771){\line( 1, 0){540}}
}%
{\color[rgb]{0,0,0}\put(5941,-781){\line( 0,-1){900}}
}%
{\color[rgb]{0,0,0}\put(2701,-781){\line( 0,-1){900}}
}%
{\color[rgb]{0,0,0}\put(6031,-1771){\line( 1, 0){540}}
}%
\put(4321,-781){\makebox(0,0)[b]{\smash{{\SetFigFont{9}{10.8}{\familydefault}{\mddefault}{\updefault}{\color[rgb]{0,0,0}$\gamma_1$}%
}}}}
\put(2701,-511){\makebox(0,0)[b]{\smash{{\SetFigFont{9}{10.8}{\familydefault}{\mddefault}{\updefault}{\color[rgb]{0,0,0}$x_1$}%
}}}}
\put(2701,-2131){\makebox(0,0)[b]{\smash{{\SetFigFont{9}{10.8}{\familydefault}{\mddefault}{\updefault}{\color[rgb]{0,0,0}$\neg x_1$}%
}}}}
\put(1981,-2131){\makebox(0,0)[b]{\smash{{\SetFigFont{9}{10.8}{\familydefault}{\mddefault}{\updefault}{\color[rgb]{0,0,0}$\neg x_1'$}%
}}}}
\put(1261,-2131){\makebox(0,0)[b]{\smash{{\SetFigFont{9}{10.8}{\familydefault}{\mddefault}{\updefault}{\color[rgb]{0,0,0}$\neg x_1''$}%
}}}}
\put(1261,-511){\makebox(0,0)[b]{\smash{{\SetFigFont{9}{10.8}{\familydefault}{\mddefault}{\updefault}{\color[rgb]{0,0,0}$x_1''$}%
}}}}
\put(1981,-511){\makebox(0,0)[b]{\smash{{\SetFigFont{9}{10.8}{\familydefault}{\mddefault}{\updefault}{\color[rgb]{0,0,0}$x_1'$}%
}}}}
\put(2341,-1636){\makebox(0,0)[b]{\smash{{\SetFigFont{9}{10.8}{\familydefault}{\mddefault}{\updefault}{\color[rgb]{0,0,0}$\gamma_2$}%
}}}}
\put(5941,-511){\makebox(0,0)[b]{\smash{{\SetFigFont{9}{10.8}{\familydefault}{\mddefault}{\updefault}{\color[rgb]{0,0,0}$x_2$}%
}}}}
\put(5941,-2041){\makebox(0,0)[b]{\smash{{\SetFigFont{9}{10.8}{\familydefault}{\mddefault}{\updefault}{\color[rgb]{0,0,0}$\neg x_2$}%
}}}}
\put(6661,-511){\makebox(0,0)[b]{\smash{{\SetFigFont{9}{10.8}{\familydefault}{\mddefault}{\updefault}{\color[rgb]{0,0,0}$x_2$}%
}}}}
\put(6661,-2041){\makebox(0,0)[b]{\smash{{\SetFigFont{9}{10.8}{\familydefault}{\mddefault}{\updefault}{\color[rgb]{0,0,0}$\neg x_2'$}%
}}}}
\put(7381,-511){\makebox(0,0)[b]{\smash{{\SetFigFont{9}{10.8}{\familydefault}{\mddefault}{\updefault}{\color[rgb]{0,0,0}$x_2$}%
}}}}
\put(7381,-2041){\makebox(0,0)[b]{\smash{{\SetFigFont{9}{10.8}{\familydefault}{\mddefault}{\updefault}{\color[rgb]{0,0,0}$\neg x_2''$}%
}}}}
\put(6301,-1636){\makebox(0,0)[b]{\smash{{\SetFigFont{9}{10.8}{\familydefault}{\mddefault}{\updefault}{\color[rgb]{0,0,0}$\gamma_3$}%
}}}}
\end{picture}%
 %

\end{center}

The operation of removal of unit clauses is also simple.
Indeed, if $l_i$ has to be removed from $F$, we can add an
edge between $l_i'$ and $l_i''$. The resulting graph has a
vertex cover of $n+m-r+1$ if and only if $F \backslash
\{l\}$ is satisfiable. Below, we see the graph obtained by
the removal of the unit clause $\gamma_2=\neg x_1$.

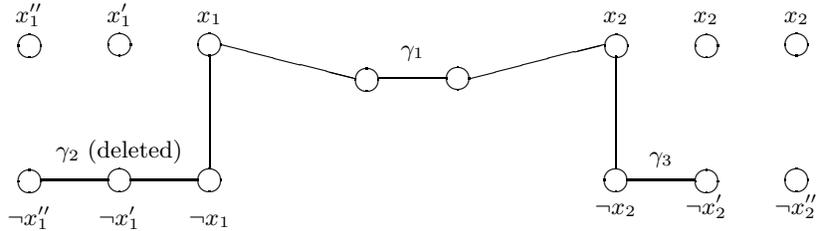
\begin{center}
\setlength{\unitlength}{3108sp}%
\begingroup\makeatletter\ifx\SetFigFont\undefined%
\gdef\SetFigFont#1#2#3#4#5{%
  \reset@font\fontsize{#1}{#2pt}%
  \fontfamily{#3}\fontseries{#4}\fontshape{#5}%
  \selectfont}%
\fi\endgroup%
\begin{picture}(6643,1834)(968,-2189)
{\color[rgb]{0,0,0}\thinlines
\put(1981,-691){\oval(180,180)}
}%
{\color[rgb]{0,0,0}\put(2701,-691){\oval(180,180)}
}%
{\color[rgb]{0,0,0}\put(1981,-1771){\oval(180,180)}
}%
{\color[rgb]{0,0,0}\put(2701,-1771){\oval(180,180)}
}%
{\color[rgb]{0,0,0}\put(3956,-971){\oval(180,180)}
}%
{\color[rgb]{0,0,0}\put(4681,-961){\oval(180,180)}
}%
{\color[rgb]{0,0,0}\put(5946,-701){\oval(180,180)}
}%
{\color[rgb]{0,0,0}\put(6666,-701){\oval(180,180)}
}%
{\color[rgb]{0,0,0}\put(5941,-1771){\oval(180,180)}
}%
{\color[rgb]{0,0,0}\put(6666,-1781){\oval(180,180)}
}%
{\color[rgb]{0,0,0}\put(1266,-701){\oval(180,180)}
}%
{\color[rgb]{0,0,0}\put(1266,-1781){\oval(180,180)}
}%
{\color[rgb]{0,0,0}\put(7381,-691){\oval(180,180)}
}%
{\color[rgb]{0,0,0}\put(7381,-1771){\oval(180,180)}
}%
{\color[rgb]{0,0,0}\put(2791,-691){\line( 4,-1){1080}}
}%
{\color[rgb]{0,0,0}\put(4051,-961){\line( 1, 0){540}}
}%
{\color[rgb]{0,0,0}\put(4771,-961){\line( 4, 1){1080}}
}%
{\color[rgb]{0,0,0}\put(2071,-1771){\line( 1, 0){540}}
}%
{\color[rgb]{0,0,0}\put(5941,-781){\line( 0,-1){900}}
}%
{\color[rgb]{0,0,0}\put(2701,-781){\line( 0,-1){900}}
}%
{\color[rgb]{0,0,0}\put(6031,-1771){\line( 1, 0){540}}
}%
{\color[rgb]{0,0,0}\put(1351,-1771){\line( 1, 0){540}}
}%
\put(4321,-781){\makebox(0,0)[b]{\smash{{\SetFigFont{9}{10.8}{\familydefault}{\mddefault}{\updefault}{\color[rgb]{0,0,0}$\gamma_1$}%
}}}}
\put(2701,-511){\makebox(0,0)[b]{\smash{{\SetFigFont{9}{10.8}{\familydefault}{\mddefault}{\updefault}{\color[rgb]{0,0,0}$x_1$}%
}}}}
\put(2701,-2131){\makebox(0,0)[b]{\smash{{\SetFigFont{9}{10.8}{\familydefault}{\mddefault}{\updefault}{\color[rgb]{0,0,0}$\neg x_1$}%
}}}}
\put(1981,-2131){\makebox(0,0)[b]{\smash{{\SetFigFont{9}{10.8}{\familydefault}{\mddefault}{\updefault}{\color[rgb]{0,0,0}$\neg x_1'$}%
}}}}
\put(1261,-2131){\makebox(0,0)[b]{\smash{{\SetFigFont{9}{10.8}{\familydefault}{\mddefault}{\updefault}{\color[rgb]{0,0,0}$\neg x_1''$}%
}}}}
\put(1261,-511){\makebox(0,0)[b]{\smash{{\SetFigFont{9}{10.8}{\familydefault}{\mddefault}{\updefault}{\color[rgb]{0,0,0}$x_1''$}%
}}}}
\put(1981,-511){\makebox(0,0)[b]{\smash{{\SetFigFont{9}{10.8}{\familydefault}{\mddefault}{\updefault}{\color[rgb]{0,0,0}$x_1'$}%
}}}}
\put(5941,-511){\makebox(0,0)[b]{\smash{{\SetFigFont{9}{10.8}{\familydefault}{\mddefault}{\updefault}{\color[rgb]{0,0,0}$x_2$}%
}}}}
\put(5941,-2041){\makebox(0,0)[b]{\smash{{\SetFigFont{9}{10.8}{\familydefault}{\mddefault}{\updefault}{\color[rgb]{0,0,0}$\neg x_2$}%
}}}}
\put(6661,-511){\makebox(0,0)[b]{\smash{{\SetFigFont{9}{10.8}{\familydefault}{\mddefault}{\updefault}{\color[rgb]{0,0,0}$x_2$}%
}}}}
\put(6661,-2041){\makebox(0,0)[b]{\smash{{\SetFigFont{9}{10.8}{\familydefault}{\mddefault}{\updefault}{\color[rgb]{0,0,0}$\neg x_2'$}%
}}}}
\put(7381,-511){\makebox(0,0)[b]{\smash{{\SetFigFont{9}{10.8}{\familydefault}{\mddefault}{\updefault}{\color[rgb]{0,0,0}$x_2$}%
}}}}
\put(7381,-2041){\makebox(0,0)[b]{\smash{{\SetFigFont{9}{10.8}{\familydefault}{\mddefault}{\updefault}{\color[rgb]{0,0,0}$\neg x_2''$}%
}}}}
\put(6301,-1636){\makebox(0,0)[b]{\smash{{\SetFigFont{9}{10.8}{\familydefault}{\mddefault}{\updefault}{\color[rgb]{0,0,0}$\gamma_3$}%
}}}}
\put(1981,-1591){\makebox(0,0)[b]{\smash{{\SetFigFont{9}{10.8}{\familydefault}{\mddefault}{\updefault}{\color[rgb]{0,0,0}$\gamma_2$ (deleted)}%
}}}}
\end{picture}%
 %

\end{center}

Assuming that $l \in F$, we have an edge between $l$ and
$l'$. By construction, either $l$ or $\neg l$ are in the
vertex cover. Now, the edge between $l$ and $l'$ enforces
$l$ to be in the cover. However, if we add an edge between
$l'$ and $l''$, and increase the size of the cover by 1, we
can add $l'$ to the cover, and then we are free to choose
$l$ or $\neg l$ to be in the optimal cover.

This proves the first part of the claim: given a graph, an
optimal vertex cover, and an edge to add, deciding whether
there exists a vertex cover for the modified graph composed
of at most $k$ nodes is \np-complete. Since removal of unit
clauses corresponds to removal of edges, we also have that
given any optimal vertex cover of the graph, if we are
allowed to add two edges, the problem remains \np-complete.

\

The third possible definition of the problem of modified
instance is that using a polynomial data structure. As for
the case of satisfiability, if a constant number of changes
is allowed, then we can create a data structure that simplifies
the problem of modified instance.

If changes are not bounded in size by a constant, we can
prove that such simplification is impossible. Note that
the reduction above cannot be used, as it only works for
addition and deletions of {\em unit clauses}. The impossibility
proof for \sat, on the other hand, uses arbitrary changes,
that is, additions and deletions of arbitrary clauses.

In this case, we show a direct proof. From the \nucnp-hardness
of the problem we prove that polynomially-sized data
structures are useless. It is possible to modify the reduction
above in such a way non-unit clauses are translated to edges
of the final graph. However, such proof would imply the
\nucnp-hardness of the vertex cover problem, and this is much
easier to be proved directly.

The proof of \nucnp-hardness is very simple: indeed, given
a set of clauses $F$ over an alphabet $X$, we consider the
graph that corresponds to the set $F'$ of all clauses of
three literals over $X$. From this graph we remove all
edges between clause nodes and literal nodes for all
clauses that are not in $F$. It is easy to see that this is
a polynomial reduction from satisfiability to vertex cover.
Moreover, since the number of nodes in the graph does not
depend on the set of clauses, but only on the number of
variables, this reduction is monotonic (the varying part of
the vertex cover problem is composed of the set of edges).

The impossibility of reducing the complexity using optimal
vertex covers of similar instances can be proved using this
compilability result. Indeed, suppose that, given a graph,
there exists a polynomially-sized piece of information that
makes polynomial the problem of vertex cover for any graph
obtained by adding a set of edges. Since any graph with $n$
nodes can be obtained by adding an appropriate number of
edges to the graph $(N,\emptyset)$, where $N$ is a set of
$n$ nodes, it follows that such piece of information makes
polynomial the problem of solving the vertex cover problem
for any graph with the given number of nodes, that is, the
problem of vertex cover would be compilable to \p, when the
set of nodes is the fixed part of the instances.

Some final comments. In the analysis of the \sat\  problem,
we gave two separate proofs for the case in which changes
may involve the creation of new variables, and the case in
which this is forbidden. For the vertex cover problem, only
one proof is needed: by construction, changes can only add
new edges, and new nodes cannot be created. Finally, note
that changes not only involve the addition of edges, but
also the bound $k$ on the number of nodes in the vertex
cover.

 %

\section{Planning}

The problems analyzed in the previous sections are rather
simple, and the results are quite unsurprising. Indeed,
it is clear that adding or removing a literal from a set
of clauses may lead to a set of clauses that is
completely different from the original one, and thus a
solution for the first instance does not help in finding
the solution of the new one. What is interesting in the
satisfiability problem is that it shows that three
possible measures of complexity can be defined when hints
are given: complexity when a specific solution is given,
complexity when an arbitrary solution is given, and
compilability (generic hint of polynomial size).

The problem of vertex cover is more interesting, as it
may be not immediately clear how solutions of similar
instances can be used. If only a solution is given, the
proofs of complexity were based on that used for
satisfiability.  The proof of \nucnp-hardness was almost
straightforward.

In this section we consider a problem of practical
interest: automated planning.  As we will show, the
proofs for the case in which an instance is given are
very similar to those of satisfiability, while the
compilability analysis is more interesting. This is due
to the fact that the fixed part of the problem is in this
case the set of actions that can be performed to achieve
the goal. While in the case of vertex cover was obvious
that the set of nodes gives not much information, in the
case of planning the set of actions is a significantly
large part of the instance.

The planning formalism we consider in this paper is
\strips.  In \strips, an instance of the planning problem
is a 4-tuple $\l {\cal P}, \O, \I, \G \r$,
where ${\cal P}$ is a set of conditions, $\O$ is the set
of operators, $\I$ is the initial state, and $\G$ is the
goal.

{\em Conditions} are facts that can be true or false
in the world of interest. A {\em state} $S$ is a set of
conditions, and represents the state of the world in a
certain time point. The conditions in $S$ are those
representing facts that are true in the world, while
those not in $S$ represent facts currently false.

The {\em initial state} is a state, thus a set of
conditions.  The {\em goal} is specified by giving a set
of conditions that should be achieved, and another set
specifying which conditions should not be made true.
Thus, a goal $\G$ is a pair $\l \M,\N \r$, where $\M$ is
the set of conditions that should be made true, while
$\N$ is the set of conditions that should be made false.

The {\em operators} are actions that can be performed to
achieve the goal.  Each operator is a 4-tuple $\l \phi,
\eta, \alpha, \beta \r$, where $\phi$, $\eta$, $\alpha$,
and $\beta$ are sets of conditions. When executed, such
an operator makes the conditions in $\alpha$ true, and
those in $\beta$ false, but only if the conditions in
$\phi$ are currently true and those in $\eta$ are
currently false. The conditions in $\phi$ and $\eta$ are
called the positive and negative {\em preconditions} of
the operator. The conditions in $\alpha$ and $\beta$ are
called the positive and negative {\em effects} or {\em
postconditions} of the operator.

Given an instance of a \strips\  planning problem $\l
{\cal P}, \O, \I, \G \r$, we define a plan for it as a
sequence of operators that, when executed in sequence
from the initial state, lead to a state where all the
conditions in $\M$ are true and all those in $\N$ are
false. More details about the definition of \strips\  can
be found in \cite{fike-nils-71} and \cite{byla-91-b}.

For the sake of simplicity, we consider a restriction of
\strips. Namely, we assume that all operators have only
positive postconditions. The problem of deciding whether
there exists a plan, under this condition, is denoted as
\plansat+, and is known to be \np-complete
\cite{byla-91-b}.

The problem of replanning arises when a plan is known,
but the scenario is modified by some external cause. In
such cases, we have a plan that is no more valid, so a
new plan is needed. Let $\l {\cal P}, \O, \I, \G \r$
be the original planning instance. We have a plan for it,
but, at some point, we discover that something we initially
believed to be true is instead false. Clearly, some
operations cannot be performed any more, so the plan may
no longer be valid. We formalize the change by assuming
that we have a modified planning instance
$\l {\cal P}, \O, \I', \G \r$. The new initial state
reflects both the fact that some actions of the initial
plan may have already been executed, and the fact that
our knowledge of the world is changed (some facts are not
true/false any more).

The first option to investigate is whether what remains
to be executed of the old plan is still a plan of the new
instance. This can be done in linear time, and a positive
outcome means that we have found the plan for the modified
instance. However, it may be that the plan is no longer
valid.

The idea behind replanning is that
the old plan may be useful in the search for the new plan,
that is, it may make the search for a new plan easier.
Formally, we are given an instance $\l {\cal P}, \O, \I, \G \r$,
one of its (optimal) plans, and a similar instance
$\l {\cal P}, \O, \I', \G \r$ with a different initial
state. We want an optimal plan for this second
instance. This is the formalization of the case in which
only a solution of the original instance is available,
like when we consider satisfiability and had a model of
the original formula as an hint.

We will prove that, even if $\I$ and $\I'$ differs from
one condition only, the problem is \np-hard, even if an
optimal plan for $\l {\cal P}, \O, \I, \G \r$ is given.

\begin{theorem}
\label{simple-fixed-plans}

Given a planning instance $\l {\cal P}, \O, \I, \G \r$,
one of its plans $P$, and a condition $a$, deciding
whether there exists a plan for $\l {\cal P}, \O, \I
\backslash \{a\}, \G \r$ is \np-complete.

\end{theorem}

\proof The problem is in \np, as we can find the new plan
just modifying the planning instance and finding the new
plan from it, and the latter problem is in \np, while
modifying the instance is polynomial.

The problem of deciding whether there
exists a plan, given the old instance, the update, and
the old plan, is \np-hard. We show that the problem of
satisfiability of a set of clauses $F$ can be reduced to
the problem of finding a plan for an updated \strips\
instance. Let $F = \{ \gamma_1 , \ldots , \gamma_k \}$ be
the set of clauses, and $X = \{ x_1 , \ldots , x_n \}$ be
its alphabet. The set of conditions of the corresponding
\strips\  instance is:\eatpar

\[
P =
\{ a \} \cup \{ t_1, \ldots , t_n \} \cup
\{ f_1 , \ldots , f_n \} \cup \{ c_1 , \ldots , c_k \}
\]

There are two conditions $t_i$ and $f_i$ for
each propositional variable $x_i$, and one condition for
each clause in the set $F$. These conditions create a
correspondence between states of the planning problem and
propositional interpretations. The conditions $t_i$ and
$f_i$ are used to represent the truth value of a variable
$x_i$, while $c_j$ holds if and only if the clause
$\gamma_j$ is true.

The initial state and the goal are specified as follows.
\eatpar

\begin{eqnarray*}
\I &=& \{ a \} \\
\G &=& \l \M,\N \r \\
\M &=& \{ c_1, \ldots , c_k \} \\
\N &=& \emptyset
\end{eqnarray*}

The set of operators is as follows.\eatpar

\[
\O =    \{ pl_1 , \ldots , pl_n \} \cup \{ nl_1 , \ldots , nl_n \} \cup
        \{ pc_1 , \ldots , pc_k \} \cup \{ nc_1 , \ldots , nc_k \} \cup
        \{ e \}
\]

\noindent where\eatpar

\begin{eqnarray*}
pl_i &=& \l \emptyset , \{ f_i,a \} , \{ t_i \} , \emptyset \r \\
nl_i &=& \l \emptyset , \{ t_i,a \} , \{ f_i \} , \emptyset \r \\
pc_i &=& \l \{ t_i \} , \emptyset , \{ c_j \} , \emptyset \r \\
nc_i &=& \l \{ f_i \} , \emptyset , \{ c_j \} , \emptyset \r \\
e  &=& \l \{ a \} , \emptyset , \{ c_1, \ldots, c_k \} , \emptyset \r
\end{eqnarray*}

These sets define the old instance of the planning
problem. The plan for the original instance is $\l e \r$.
It can be easily verified that this is a plan for the
above instance (namely, it is the only minimal and
irredundant one).

The considered update is the deletion of the condition
$a$ from the initial state. Since there is no operator
that has $a$ among its postconditions, it follows that
$a$ does not hold in any state reached from the initial
state. As a result, the operator $e$ have no effect, thus
it can be deleted from the instance, as well as $a$.

The obtained instance is equivalent to that given in
\cite{byla-91-b} to prove that \plansat+ is \np-complete.
What is proved in Bylander's paper is that there exists a
plan for the considered instance if and only if the set
of clauses $F$ is satisfiable.~\qed

Taking a closer look to the instance considered in the
proof, we notice that the old plan does not help because
the result of the change is an instance on which the only
action of the old plan cannot be executed at all. This
means that the old plan is now completely useless.

As for the satisfiability and the vertex cover problems,
a possible objection is that the hardness result proves
is only that there are useless plans, not that all plans
are useless. The question of whether there exists plans
that help in solving similar planning instance can be
answered with the very same proof of the previous
theorem.

\begin{theorem}\label{simple-any-plan}

Given any minimal or any irredundant plan for a planning
instance, finding a plan for an updated instance is
\np-complete, even if the update is the deletion of a
condition from the initial state.

\end{theorem}

\proof In the planning instance used in proof of
Theorem~\ref{simple-fixed-plans}, there is exactly one
irredundant plan, which is also the only minimal plan:
the sequence $\l e \r$. As a result, given any set of
propositional clauses $F$, we build the planning instance
and the update as in the proof of
Theorem~\ref{simple-fixed-plans}. Now, given any
irredundant (or minimal) plan for this instance, we
proved that $F$ is satisfiable if and only if there
exists a plan for the updated instance.~\qed

To complete the analysis, only one step is missing: the
case in which a polynomially-sized data structure is
avaliable. This encodes also the case in which some
intermediate result, found during the search for the
previous plan, is saved for later use with modified
planning instances. This data structure can only
depend on the original instance.

If a constant number of changes is allowed (that is,
the initial state can only be modified for a constant
number of conditions that are either added or deleted),
then a data structure that helps finding a plan for
the modified problem is easy to define.

We therefore consider the case in which the initial
states of the original and the new instance can be
different. Note, however, that the two instances have
the same conditions, operators, and goals. This is
a constraint over the possible change, and loosely
corresponds to the case of satisfiability when no
new variables can be added. Indeed, the impossibility
proof make use of compilability classes.

\begin{theorem}

If there exists a polynomial data structure such that
deciding the existence of plans for updated instances is
polynomial, then \S{2}=\P{2}, \ie, the polynomial
hierarchy collapses.

\end{theorem}

\proof We use a result on non-compilability of planning
given in \cite{libe-98}. The result is: given a set of
conditions ${\cal P}$ and operators $\O$, there is no
polynomial data structure that allows the solving of a
generic instance $\l {\cal P}, \O, \I, \G \r$ in
polynomial time. This implies that it is not possible to
{\em compile} ${\cal P}$ and $\O$ in such a way the
solving for any initial state and goal is polynomial.

We prove that the same holds when the part of the instance
that is compiled is composed by ${\cal P}$, $\O$, and $\G$,
that is, only the initial state is not compiled. This is
shown by showing a \nuc-reduction from the problem of
planning (in which ${\cal P}$ and $\O$ are the fixed part
of the instance) to the problem of planning (in which only
$\I$ is the varying part of the input).

Let $\l {\cal P}, \O, \I, \G \r$ be an instance of the
original planning problem, where $\G=\l \N,\M \r$.
We derive a new instance
$\l {\cal P}\cup \{g\}, \O\cup\{o\}, \I, \{\{g\},\emptyset\} \r$
that has a plan if and only if the original one has. The
new instance has one more condition $g$, used to formalize
the reachability of the goal in the original one, and a new
action $o$ used for the same aim, and defined as
$o=\l \N, \M, g, \emptyset \r$. In words, this action can
be used, whenever the goal of the original problem is
satisfied, to make $g$ true. As a result, a plan for the
original problem exists if and only if the new instance
admit a plan. In the new instance, the goal is constant
(does not depend at all from the original instance). Therefore,
its membership to the fixed or varying part of the input is
irrelevant. This proves that the problem of planning, where
the initial state is the only varying part of the input,
is \nucnp-complete.

We use this result for proving that it is not possible to
find, given a planning instance, a polynomially-sized data
structure that allows solving similar problems.
Suppose that such a structure exists. Then, it
would be possible to compile ${\cal P}$, $\O$, and $\G$ just
by finding such data structure for $\l {\cal P}, \O, \emptyset,
\G \r$. Since any \strips\
instance $\l {\cal P}, \O, \I, \G \r$ can be obtained by
updating $\l {\cal P}, \O, \emptyset, \G \r$, the data
structure makes any such instance solvable in polynomial time,
and this contradicts the result given above.~\qed

 %

\section{Conclusions}
\label{conclusions}

This paper report on the study of how the complexity of
a problem changes when a solution of a similar instance
is given. The most interesting result is that three possible
characterizations can be given for this problem. The simplest
one is simply the one in which we assume that a given
solution is part of the input. While the hardness proof
of this case is quite straightforward, it only proves that
there are solutions that do not help in solving similar
instances. We therefore considered the more interesting case
in which any solution is considered. Finally, we considered
the case in which the available hint is not only composed
of a solution of a similar instance, but may be any
polynomially-sized data structure depending on the similar
instance only. This formalizes, for example, the case in
which some information, derived from the search of the
solution of the similar instance, is saved for use on
changed instances.

The hardness results of this paper are proven in two ways:
either we provide a direct proof of the claim (for example,
that satisfiability is not simplified, if a model of a similar
instance is given), or obtained by employing a special kind
of reductions, in which small changes to the original instance
corresponds to small changes of the resulting one. Finally,
for the case in which polynomially-sized data structures are
available, and there are constraints on the kind of possible
changes (like, new variables cannot be introduced), the
compilability classes have been used to prove impossibility
results.

The case in which a polynomially-sized data structure comes
from generalizing the case in which solving an instance of
a problem we find some intermediate results that may be useful
for solving similar instance. On the other hand, this may
look as an over-generalization. In other words, not all
polynomially-sized data structures are reasonable partial
results of algorithms. In particular, if the changes are small,
we used some data structures containing the solution for any
possible modified instance. It is clearly unlikely that an
algorithm produces such table by accident. However, some facts
are to be considered:

\begin{enumerate}

\item while solving an instance, we are aware that it may
be later modified; therefore, such table may be the actual
result of an algorithm that is designed taking this into account;

\item the restricted case in which the additional information
is something found during the solution of the original
instance is hard to analyze; indeed, the kind of additional
information depends on the specific algorithm employed;

\item in some cases, the first instance may be solved
without finding any additional information; in the case
of satisfiability, the DPLL algorithm may find the model
immediately, by making (by accident) the
right choice of the first sign of the literal to expand at
each branching point; a local search algorithm may find
the model as it is the first one it tries; in both cases,
no other information is available;

Note that such cases are not that unlikely: for
example, DPLL may find a model of $a \vee G$ because $a$ is
the literal occurring most often, and is chosen as the first
branching literal: if the positive sign is chosen first, then
a model is found immediately; for local search,
the formula $a \vee G$ is satisfied by all models with $a$,
so we have at least 50\% probability that the first model
tried by the algorithm (chosen at random)
contains $a$; in both cases, we have no additional information
that can help us if $\neg a$ is added to the set.

Finally, note that proof complexity can hardly be used in
this context, as it is mainly oriented to proving unsatisfiability,
and mainly when exponentially large proofs are needed.

\end{enumerate}

 %

\bibliographystyle{alpha}

\begin{thebibliography}{CDLS02}

\bibitem[Byl91]{byla-91-b}
T.~Bylander.
\newblock Complexity results for planning.
\newblock In {\em Proceedings of the Twelfth International Joint Conference on
  Artificial Intelligence (IJCAI'91)}, pages 274--279, 1991.

\bibitem[CDLS96]{cado-etal-96}
M.~Cadoli, F.~M. Donini, P.~Liberatore, and M.~Schaerf.
\newblock Feasibility and unfeasibility of off-line processing.
\newblock In {\em Proceedings of the Fourth Israeli Symposium on Theory of
  Computing and Systems (ISTCS'96)}, pages 100--109. {IEEE} Computer Society
  Press, 1996.

\bibitem[CDLS02]{cado-etal-02}
M.~Cadoli, F.~Donini, P.~Liberatore, and M.~Schaerf.
\newblock Preprocessing of intractable problems.
\newblock {\em Information and Computation}, 176(2):89--120, 2002.

\bibitem[FN71]{fike-nils-71}
R.~Fikes and N.~Nilsson.
\newblock {STRIPS}: a new approach to the application of theorem proving to
  problem solving.
\newblock {\em Artificial Intelligence}, 2:189--208, 1971.

\bibitem[Lib98]{libe-98}
P.~Liberatore.
\newblock On the compilability of diagnosis, planning, reasoning about actions,
  belief revision, etc.
\newblock In {\em Proceedings of the Sixth International Conference on
  Principles of Knowledge Representation and Reasoning (KR'98)}, pages
  144--155, 1998.

\end{thebibliography}

\end{document}